\documentclass[english,11pt,superscriptaddress]{article}
\pdfoutput=1
\usepackage{lmodern}

\usepackage[T1]{fontenc}
\usepackage[utf8]{inputenc}
\usepackage{geometry}
\usepackage{array}
\usepackage{amsmath}
\usepackage{graphicx}
\usepackage{float}
\usepackage{color}
\usepackage{hyperref}
\usepackage{framed}
\usepackage{changes}
\usepackage{float}
\usepackage{array}
\usepackage{tikz}
\usetikzlibrary{backgrounds}
\usepackage{comment}

\makeatletter

\providecommand{\tabularnewline}{\\}

\usepackage[winfonts,UTF8]{ctex}
\usepackage{slashed}
\usepackage{hyperref}
\usepackage{bbding}
\usepackage{graphicx}
\usepackage{subfigure}
\usepackage[numbers,sort&compress]{natbib}
\date{}

\hypersetup{
colorlinks=false,
linkcolor=black
}

\makeatother

\usepackage{babel}

\def\be{\begin{equation}}
\def\ee{\end{equation}}
\def\sR{\mathsf{R}}

\def \mc {\mathcal}
\begin{document}
\title{\textbf{The Fate of Instability of  de Sitter Black Holes at Large $D$}}
\author{Peng-Cheng Li$^{1,2}$\footnote{lipch2019@pku.edu.cn}~,
Cheng-Yong Zhang$^{3}$\footnote{zhangcy@email.jnu.edu.cn}~,
and
Bin Chen$^{1,4,5}$\footnote{bchen01@pku.edu.cn}}
\date{}

\maketitle

\vspace{-10mm}

\begin{center}
{\it
$^1$Department of Physics and State Key Laboratory of Nuclear
Physics and Technology, Peking University, No.5 Yiheyuan Rd, Beijing
100871, P.R. China\\\vspace{1mm}

$^2$School of Physics and Astronomy, Sun Yat-sen University, Zhuhai 519082, China\\\vspace{1mm}

$^3$Department of Physics and Siyuan Laboratory, Jinan University, Guangzhou 510632, China\\\vspace{1mm}

$^4$Center for High Energy Physics, Peking University,
No.5 Yiheyuan Rd, Beijing 100871, P. R. China\\\vspace{1mm}

$^5$ Collaborative Innovation Center of Quantum Matter,
No.5 Yiheyuan Rd, Beijing 100871, P. R. China
}
\end{center}

\vspace{8mm}

\begin{abstract}
We study non-linearly the gravitational instabilities of the Reissner-Nordstrom-de Sitter and the Gauss-Bonnet-de Sitter black holes by using the large $D$ expansion method. In both cases, the thresholds of the instability are found to be consistent with the linear analysis, and  on the thresholds the evolutions of the black holes under the perturbations settle down to stationary lumpy solutions. However, the solutions in the unstable region are  highly time-dependent, and resemble the fully localized black spots and black ring with $S^{D-2}$ and $S^1\times S^{D-3}$ topologies, respectively.
Our study indicates the possible transition between the lumpy black holes and the localized black holes in higher dimensions.
\end{abstract}

\maketitle

\newpage

\section{Introduction}
  One important aspect of a black hole is its stability under gravitational perturbations. 
  In contrast to the $D=4$ case, the black holes in higher dimensions can have various topologies  and could be unstable under gravitational perturbations. This was first illustrated by Gregory and Laflamme (GL) for the black strings (branes) \cite{Gregory1993}. Since then, a lot of black objects with distinct topologies  in higher dimensions have been discovered, and similar instabilities have been demonstrated \cite{Kol2006,Harmark2007,Emparan2003}. Compared with the linear dynamics, determining the final state of the unstable black objects is more
 intriguing, as which may lead to violation of the Weak Cosmic Censorship and topology-changing
transitions of black hole horizons. A well-known example is the non-linear evolution of the GL instability of the five dimensional black string \cite{Lehner2010}.

Recently, a new type of dynamical instability was discovered for the higher dimensional Reissner-Nordstrom-de
Sitter (RN-dS) and Gauss-Bonnet-de Sitter (GB-dS) black holes \cite{Konoplya2008,Konoplya:2013sba, Cuyubamba2016}. The RN-dS black hole becomes  gravitationally unstable for large values of the electric charge and cosmological constant. The GB-dS black hole is unstable as well when the cosmological constant is sufficiently large. In both cases the instability is triggered by  a positive cosmological constant. Such instability  is called  ``$\Lambda$ instability''. One interesting feature of the ``$\Lambda$ instability'' is that such instability only happens with an electromagnetic field or a GB term. Moreover, it was noticed that at the threshold point of the instability the shape of the  RN-dS black hole is slightly deformed. This fact suggests that the unstable RN-dS black hole could either split into two black holes or transform into a black ring. However, due to the limitation of the linear analysis, the final stage of the evolution of the unstable black holes has not been settled yet, though some non-linear efforts have been made from the viewpoint of gravitational collapse \cite{Zhang2015}.

The goal of this work is to better understand the ``$\Lambda$ instability'' of the two kinds of black holes  by using the large $D$ expansion method \cite{Emparan2013}. Taking the large dimension limit, the non-trivial dynamics of black holes is localized at the near region of the horizon, such that a full non-linear effective theory can be formulated \cite{Emparan:2014aba,Emparan:2015hwa,Bhattacharyya:2015dva,Suzuki:2015iha,Emparan:2015gva,Bhattacharyya:2015fdk,Dandekar:2016fvw, Bhattacharyya:2017hpj, Bhattacharyya:2018szu,Kundu:2018dvx}. The large $D$ studies of the RN-dS and GB-dS black holes were initiated in \cite{Tanabe2015} and \cite{Chen2017}. It was shown there that at large $D$ the dynamical equations for the RN-dS and GB-dS black holes are reduced to much simpler 
effective equations such that the ``$\Lambda$ instability'' can be demonstrated by performing linear analysis for these equations.
In this paper, we take a further step to perform a full non-linear study of the instability, including the analysis of the static solutions and the time evolution of the dynamical equations. {We will analyze the impacts of the electric charge and the GB constant on the instability.} More importantly we will study the possible final fates of the ``$\Lambda$ instability''. The non-linear stabilities of the RN and GB black holes in the asymptotically flat and AdS spacetimes will be addressed as well.

The paper is organized as follows. In section \ref{sec:Eff}, we introduce the effective equations of gravity at large $D$  and the static solutions by the name of  lumpy black holes. In section \ref{sec:Nonl}, we study in detail the time evolution of the RN-dS   and GB-dS black holes under the perturbations at large $D$, both numerically and analytically. In section \ref{sec:Conclusion}, we summarize our studies and show some limitations about the present work.

\section{\label{sec:Eff}Lumpy black holes}
We begin with  a brief review of the large $D$ expansions of the RN-dS and GB-dS black holes in \cite{Tanabe2015} and \cite{Chen2017}.  We focus on the large $D$ RN-dS black hole first. The metric for a  spherically symmetric, dynamical charged black hole in the de Sitter spacetime can be  written in terms of the ingoing Eddington-Finkelstein
coordinates as
\be
ds^2=-Adt^2+2u_tdtdr-2C_zdtdz+r^2Gdz^2+r^2\sin^2zd\Omega_n^2,
\ee
where $z$ is a coordinate on  the sphere $S^{D-2}$ and $n=D-3$. The Maxwell field is
\be
A_\mu dx^\mu=A_tdt+A_zdz.
\ee
Since at large $D$, the radial derivative becomes dominant such that the radial dependence of the Einstein-Maxwell equations can be integrated out firstly, which
gives
\be
A=1-\frac{{\bf m}}{\sR}+\frac{{\bf q}^2}{\sR^2},\quad u_t=1, \quad G=1,
\ee
\be
C_z=\frac{1}{n}\left(\frac{{\bf p}}{\sR}-\frac{{\bf p}{\bf q}^2}{\sR^2}\right),\quad A_t=\frac{\sqrt{2}{\bf q}}{\sR},\quad A_z=-\frac{\sqrt{2}}{n}\frac{{\bf p}{\bf q}}{{\bf m}\sR},
\ee
where we have used a new radial coordinate $\sR=r^n$. The functions ${\bf m}(t,z)$, ${\bf q}(t,z)$ and ${\bf p}(t,z)$ represent the mass, electric charge and momentum distributions along the horizon, respectively.
Then the remaining Einstein-Maxwell equations are reduced to the following effective equations
\begin{align}
0= & \partial_{t}{\bf q}+\cot z\partial_{z}{\bf q}+\cot z\frac{{\bf q}}{{\bf m}}{\bf p},\label{eq:effQ}\\
0= & \partial_{t}{\bf m}+\cot z\partial_{z}{\bf m}+\cot z\,{\bf p},\label{eq:effM}\\
0= & \partial_{t}{\bf p}-\frac{\sqrt{{\bf m}^{2}-4{\bf q}^{2}}\cot z}{ {\bf m}}\partial_{z}{\bf p}
 +\left(1-\hat{\Lambda}+\frac{(\sqrt{{\bf m}^{2}-4{\bf q}^{2}}-{\bf m}){\bf p}}{{\bf m}^{2}}\cot z\right)\partial_{z} {\bf m} \label{eq:effP}\\
 & -\left(1-\frac{{\bf p}}{{\bf m}}\cot z-\frac{\sqrt{{\bf m}^{2}-4{\bf q}^{2}}}{{\bf m}}\cot^{2}z\right){\bf p},\nonumber
\end{align}
where $\hat{\Lambda}$ is the cosmological constant, and its value is restricted to $\hat{\Lambda}\leq1$.
Note that $\hat{\Lambda}$ is related to the  cosmological constant in the action by $\Lambda=(D-1)(D-2)\hat{\Lambda}/2$. These equations encode the non-linear dynamics of the RN-dS black holes.

\begin{figure}[h]
\begin{centering}
\begin{tabular}{>{\centering}p{6cm}>{\centering}p{6cm}}
\includegraphics[scale=0.2]{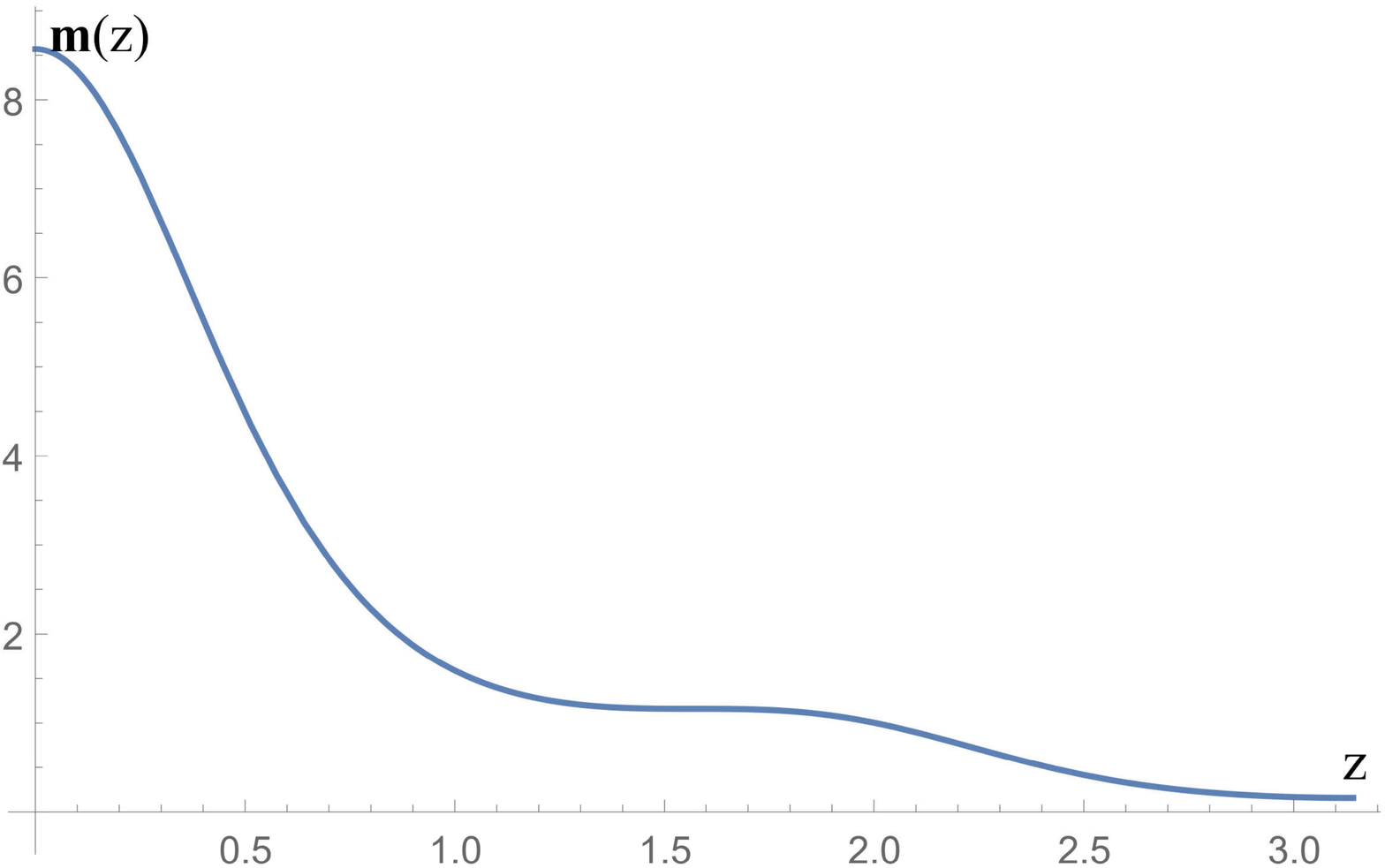}

(a) $P_1>0, P_2 =3$  &
\includegraphics[scale=0.2]{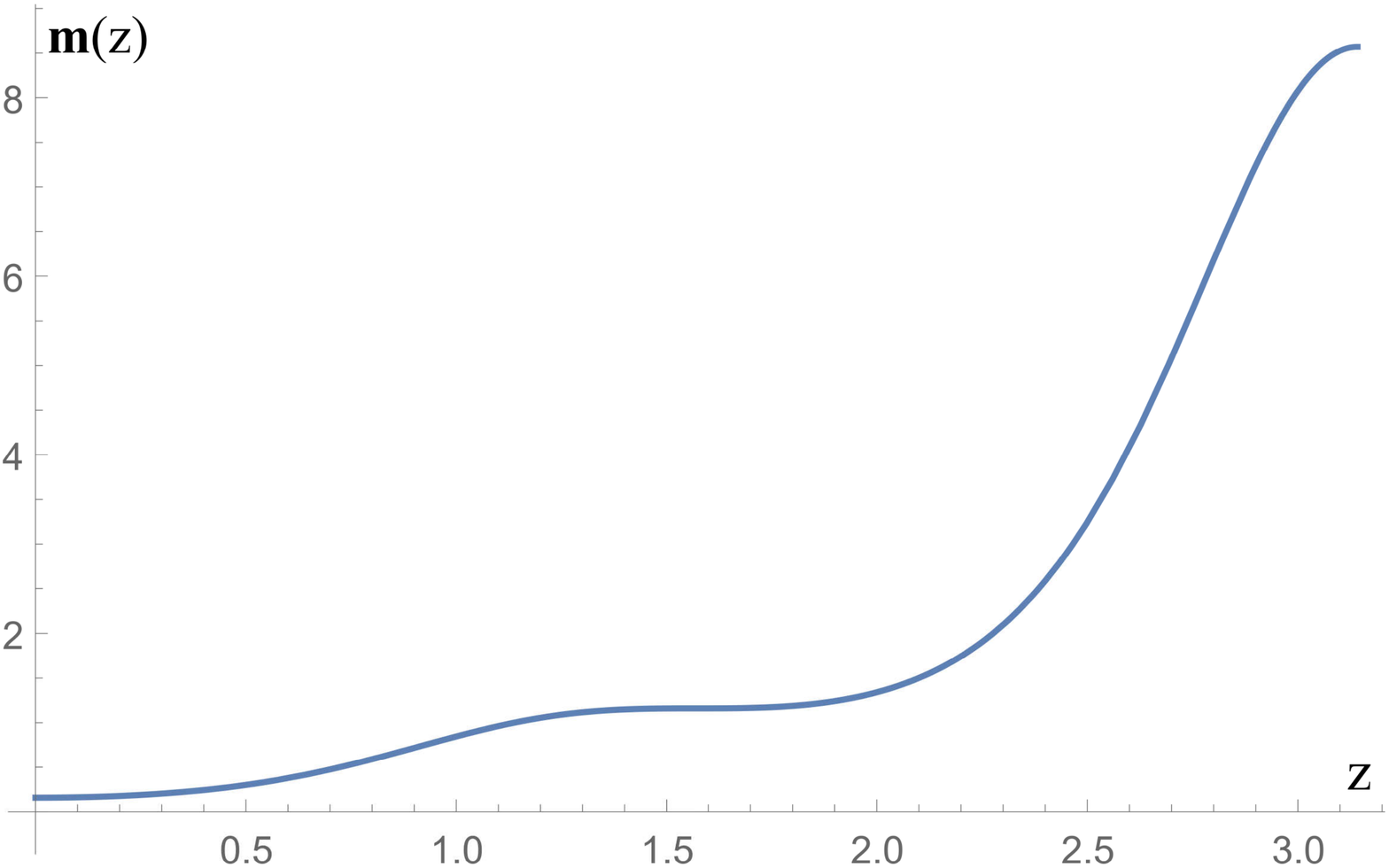}

(b) $P_1<0, P_2 =3$\tabularnewline
\includegraphics[scale=0.2]{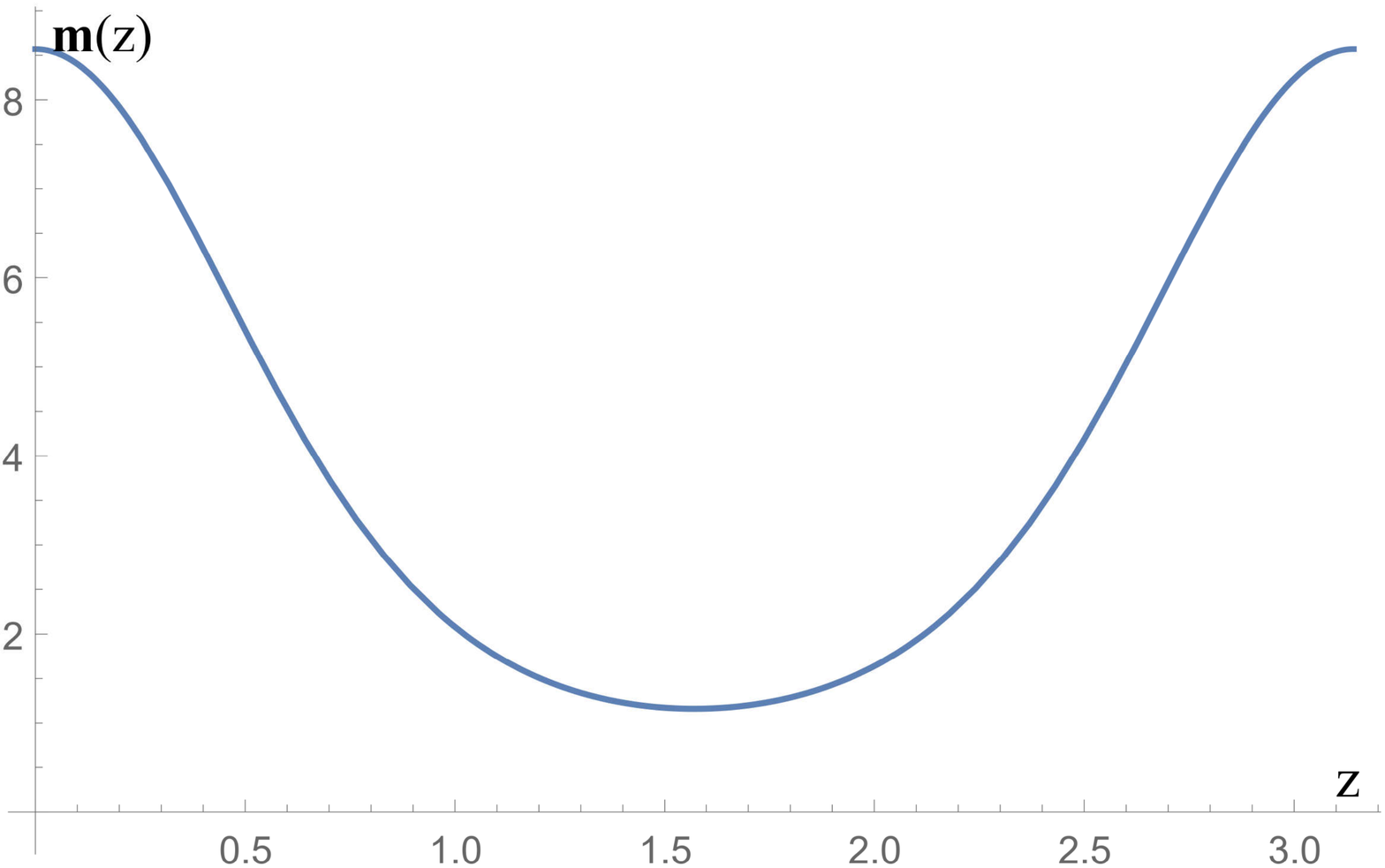}

(c)
$P_1>0, P_2=2$ &
\includegraphics[scale=0.2]{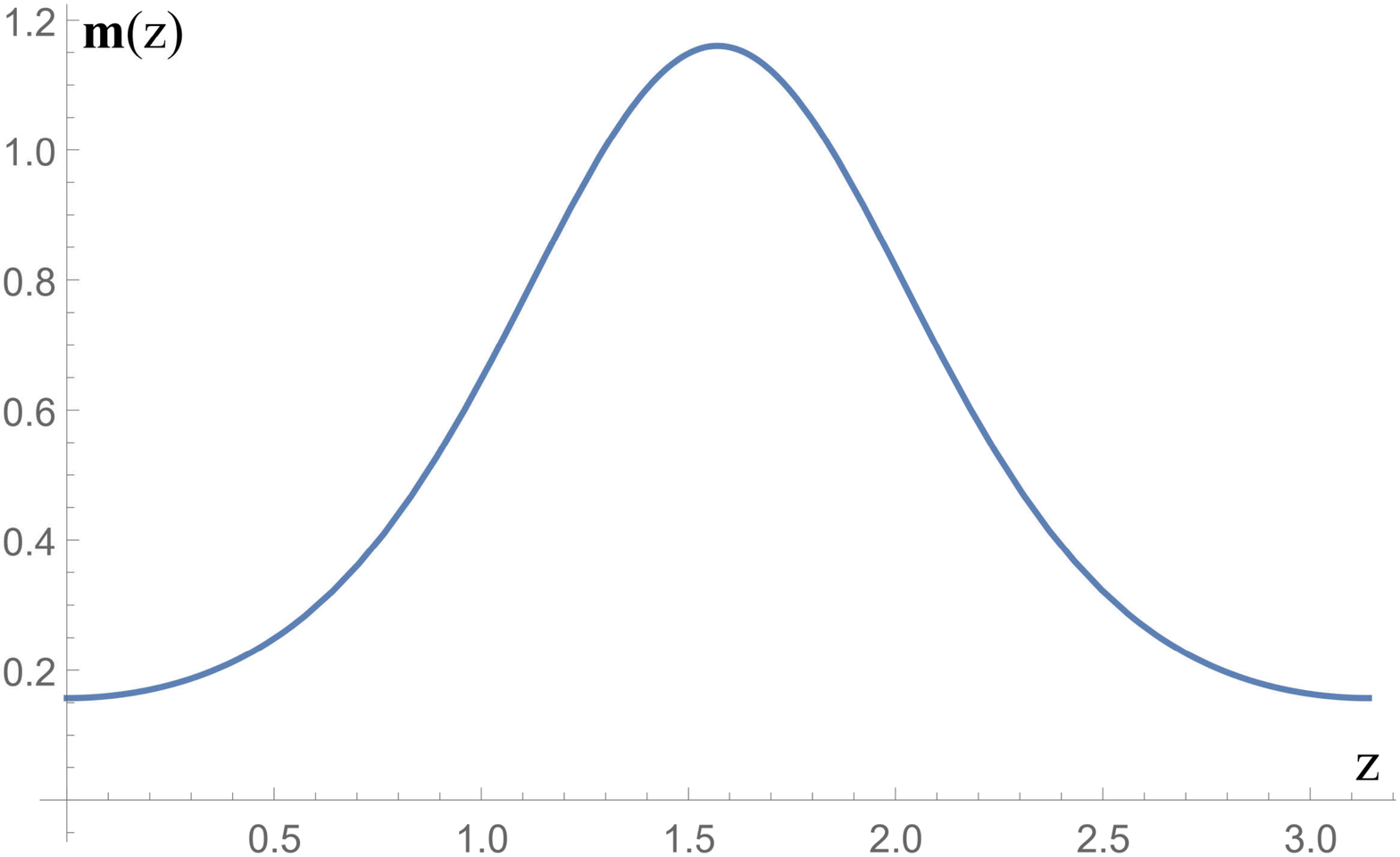}

(d) $P_1<0, P_2=2$\tabularnewline
\end{tabular}
\par\end{centering}
\caption{\label{fig:RNlumpyBHs} Branches of lumpy black holes. The horizontal
axis is $z$ and the vertical axis is the mass density ${\bf m}(z)$. In these panels, we take $Q^2=4/25$, $P_1=\pm2$ and $P_2=2$ or 3 for illustration. Note that the measures on the vertical axes in the figures are not uniformized in order to show the symmetry more clearly. }

\end{figure}

The static and uniform  solution  of the above equations is $z$-independent with
${\bf p}=0$, ${\bf m}=1+Q^{2}$ and ${\bf q}=Q$,
where  the horizon radius is set to unity and the extreme case corresponds to $Q^2=1$.
As shown by the linear analysis in \cite{Tanabe2015},  there is a zero mode at the edge of the instability, which generates a new family non-uniform solutions being called lumpy back holes. These lumpy black holes belong to the static solutions of the effective equations, i.e.
\begin{align}
{\bf m}(z)&=(1+Q^{2})e^{P_1(\cos z)^{P_2} },\label{eq:Deformed}\\
{\bf p}(z)&={\bf m}'(z), \nonumber\\
 {\bf q}(z)&=\frac{Q}{1+Q^{2}}{\bf m}(z),\nonumber
\end{align}
where
$
P_{2}=1+
\hat{\Lambda}\frac{(1+Q^{2})}{1-Q^{2}}>1,\label{eq:DeformedP2}
$
and $P_{1}$ is an integration constant describing the deviation from the spherical symmetry.
%
For general $Q$ and $\hat{\Lambda}$, the solution (\ref{eq:Deformed}) may not be regular at $z\geq\pi/2$ unless $P_2$ is an positive integer \footnote{The quantization condition for $P_2$ emerges not only from
$z=\pi/2$, but also from the region $z\in(\pi/2,\pi)$ since $(\cos z)^{P_{2}}$
would be a complex number if $P_{2}$ is not an integer. Further,
$P_{2}$ cannot be an negative integer because $(\cos z)^{P_{2}}$
would be infinity when $z\to\pi/2$.}.
 The linear analysis finds that under the perturbation the black hole settles down to a lumpy black hole which is  static when $Q$ and $\hat{\Lambda}$ take the values such that $P_2=\ell$, with $\ell$ the harmonics on $S^{D-2}$ \cite{Tanabe2015}. These lumpy black holes break the $SO(D-1)$ rotation symmetry down to $SO(D-2)$ but share the same horizon topology $S^{D-2}$ with the uniform black hole.
 Depending on   whether $P_1$ is positive or negative and whether $P_2$ is odd or even, we find  three branches for the lumpy black holes,  as shown in Fig. \ref{fig:RNlumpyBHs}. In Fig. \ref{fig:RNlumpyBHs} we only show the case that $P_1$ takes the same value for both regions $z<\pi/2 $ and $z>\pi/2 $, and so when $P_2$ is odd, the cases $P_1>0$ and $P_1<0$ are the same under $z\to \pi -z$. In fact $P_1$ can be different in the two regions, which is not in conflict with the quantization for $P_2$.
%
%

From the thermodynamics we know that in the microcanonical ensemble, the preferred phase is the one with a larger entropy.
 At the leading order of the $1/D$ expansion, the mass and entropy of the lumpy black holes
are given by
\begin{align}
\mathcal{M}&=\frac{\Omega_n}{16\pi}n\int_0^\pi{\bf m} \sin^{D-3}z\, dz,\\
\mathcal{S}&=\frac{\Omega_n}{4}\int_0^\pi \frac{{\bf m}+\sqrt{{\bf m}^{2}-4{\bf q}^{2}}}{2} \sin^{D-3}z\, dz.
\end{align}
The factor $\sin^{D-3} z$ in the integral means that  the dominant contribution to the integration comes from the region $z\simeq\pi/2$ at large $D$. As a result, all the masses and entropies of different solutions are equal. The difference of the entropy might be manifested by taking the  $1/D$ correction into account. But this is not an easy task and we will not pursue it in this paper. 



\section{\label{sec:Nonl}Time evolution}

In fact, the time evolution of the black hole under the perturbations  can  tell which phase is preferred. For example, though the uniform and non-uniform black strings have the same entropy at the leading order of the $1/D$ expansion,  the time evolution of the unstable black strings clearly shows that the non-uniform solutions are the stable end point. This is in accord with the entropy argument for the static solutions  at the next-to-leading order of the $1/D$ expansion \cite{Suzuki1506,Emparan:2015gva,Rozali1607,Emparan:2018bmi}. In this section, we will study the time evolution of the RN-dS and GB-dS black holes under the perturbations.

\subsection{RN-dS black hole}

We study the time evolution of the large $D$ effective equations of the RN-dS black hole now.
From (\ref{eq:effM}), it is obvious that ${\bf m}$ is invariant
at $z=\pi/2$, thus we take the boundary condition
  ${\bf m}(t,\frac{\pi}{2})=1+Q^{2}$.
The fact that the total momentum should be zero $\int {\bf p}(t,z)(\sin z)^{D-3} dz=0$  implies the boundary condition ${\bf p}(t,\frac{\pi}{2})=0$ in the large $D$ limit.
For charge density we take ${\bf q}(t,\frac{\pi}{2})=\frac{Q}{1+Q^{2}}{\bf m}(t,\frac{\pi}{2})$
which is consistent with the static solution.
The initial conditions are taken to be
\begin{align}
{\bf m}(0,z)= &\, 1+Q^{2}+\delta {\bf m}\left(z\right)-\delta {\bf m}\left(\frac{\pi}{2}\right),\\
{\bf q}(0,z)= & \,\frac{Q}{1+Q^{2}}{\bf m}(0,z), \quad
 {\bf p}(0,z)=0.
\end{align}
Here
\be
\delta{\bf m}\left(z\right)=\sum_{i=1}^{n_{m}}a_{i}\cos(b_{i}z+c_{i}),\ee
in which $n_{m}$ is the number of perturbation modes and it takes value 20 here, $a_{i}$ are
random numbers from 0 to $10^{-3}$, $b_{i}$ from 0 to $32$ and
$c_{i}$ from 0 to $2\pi$. Under these conditions,  the effective equations can be handled by using the {\em Mathematica} function {\bf NDSolve}.
\begin{figure}
\begin{centering}
\begin{tabular}{>{\centering}p{6cm}>{\centering}p{6cm}}
\includegraphics[scale=0.25]{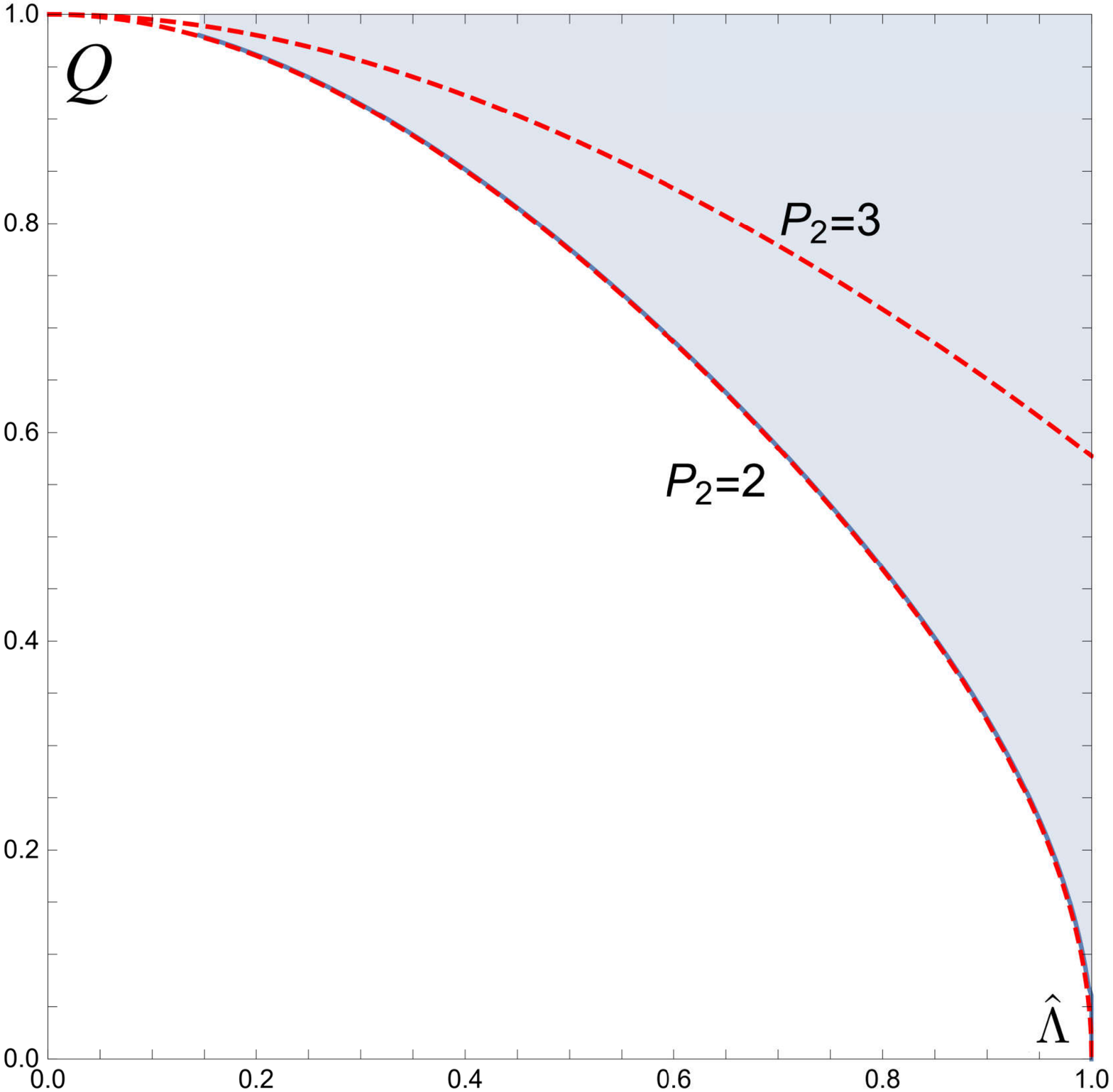} & \includegraphics[scale=0.25]{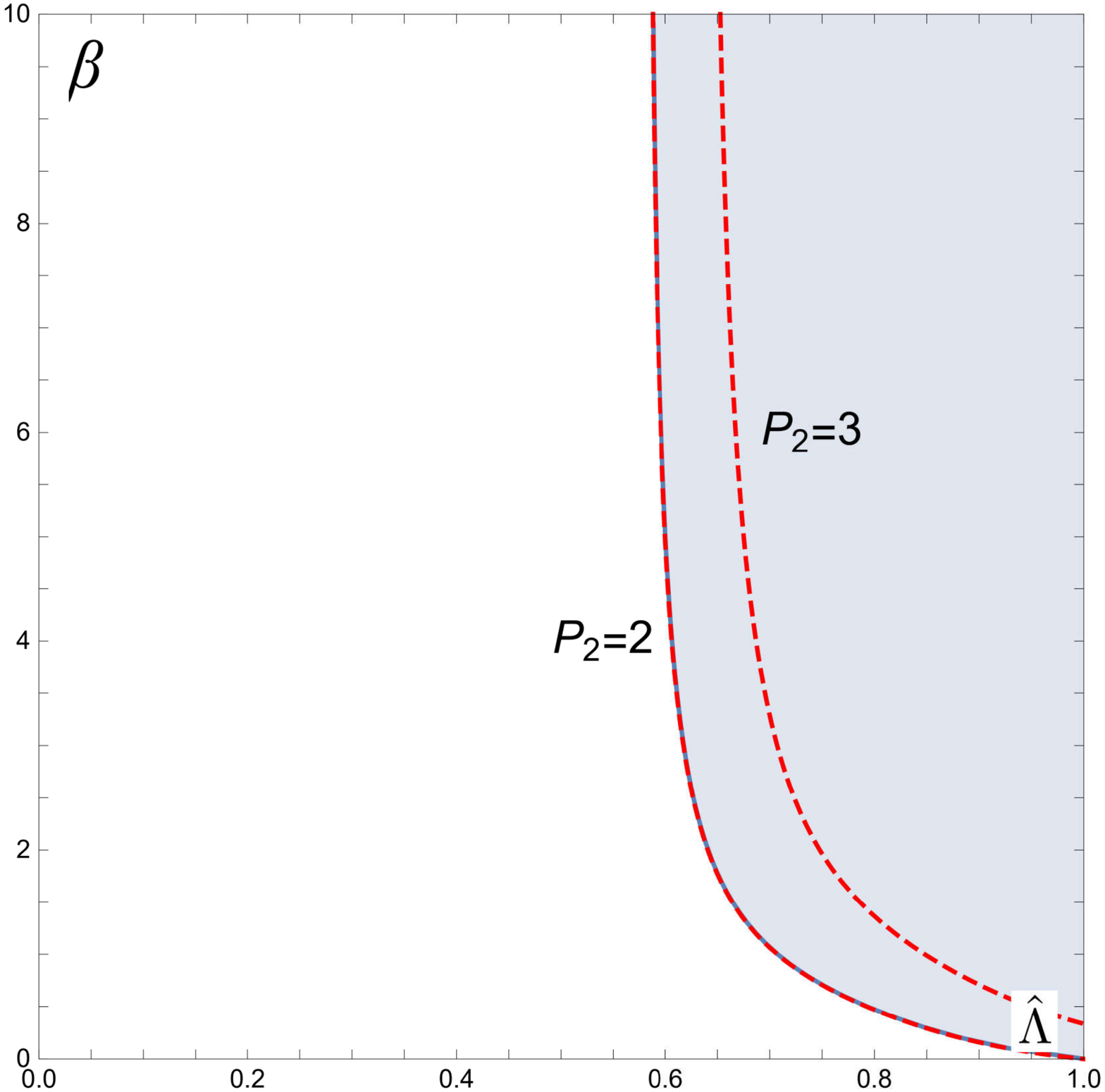}\tabularnewline
\end{tabular}
\par\end{centering}
\caption{\label{fig:RNGBdSUnstableRegion}In the parameter spaces, the unstable regions are in blue  for the RN-dS
(left) and GB-dS (right) black holes from the non-linear
evolution, the threshold is along the line $P_2=2$. The horizontal axis is $\hat{\Lambda}$ and the vertical
axis is $Q$ (left) or $\beta$ (right). The red lines
represent the conditions that the linear
perturbation  is static \cite{Tanabe2015,Chen2017}. }
\end{figure}

The threshold of the parameters $(\hat{\Lambda}, Q)$ for the instability is shown in the left panel of Fig. \ref{fig:RNGBdSUnstableRegion}.  Note that the instability occurs only when both a positive cosmological constant and an electric charge are present.
{The threshold coincides with the onset of the linear instability when $P_{2}\equiv 1+\hat{\Lambda}\frac{(1+Q^{2})}{1-Q^{2}} =2$.
The black holes with $P_{2}>2$ are unstable. So in principle there is not static solution when $P_2>2$. This result is consistent with the linear analysis. According to the linear analysis, the quasinormal mode depends on $(\ell, \hat{\Lambda}, Q)$. When $P_{2}=N\geq2$, the mode  $\ell=N$ is static, but the modes $\ell<N$ are unstable and the modes $\ell>N$ dispates (here $N$ is a positive integer). When different modes are superposed, the only possible static solution is that with both $P_2=2$ and $\ell=2$. This point will be explained clearer in the following with the help of the analytical solution.}

In the stable region, the perturbation dissipates and leaves a uniform black hole just the same as the RN-dS black hole before the perturbation.
\begin{figure}[h]
\begin{centering}
\begin{tabular}{>{\centering}p{6cm}>{\centering}p{6cm}}
\includegraphics[scale=0.22]{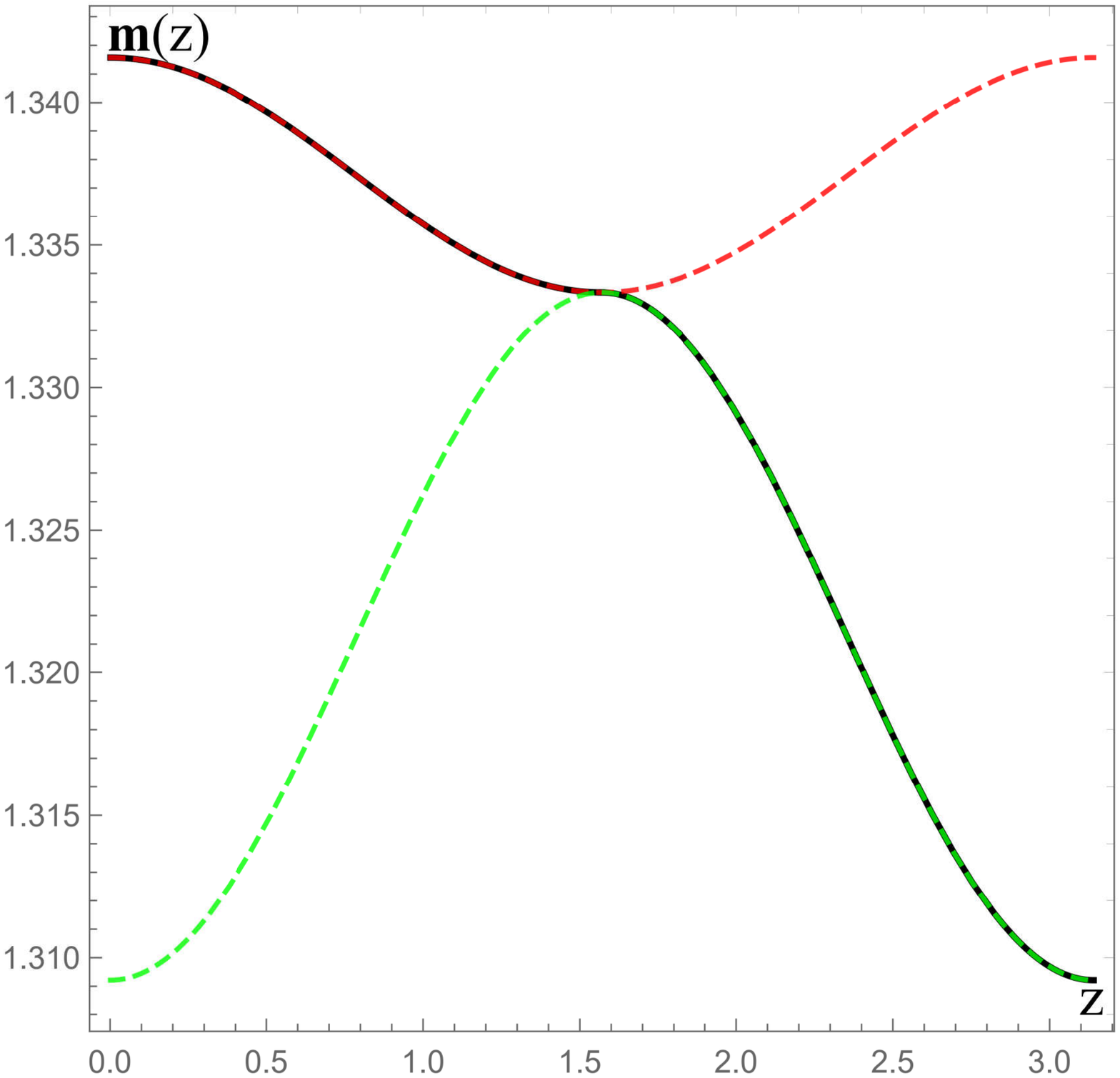}

(a) ${\bf m}'(0,\frac{\pi}{2})>0$ &

\includegraphics[scale=0.22]{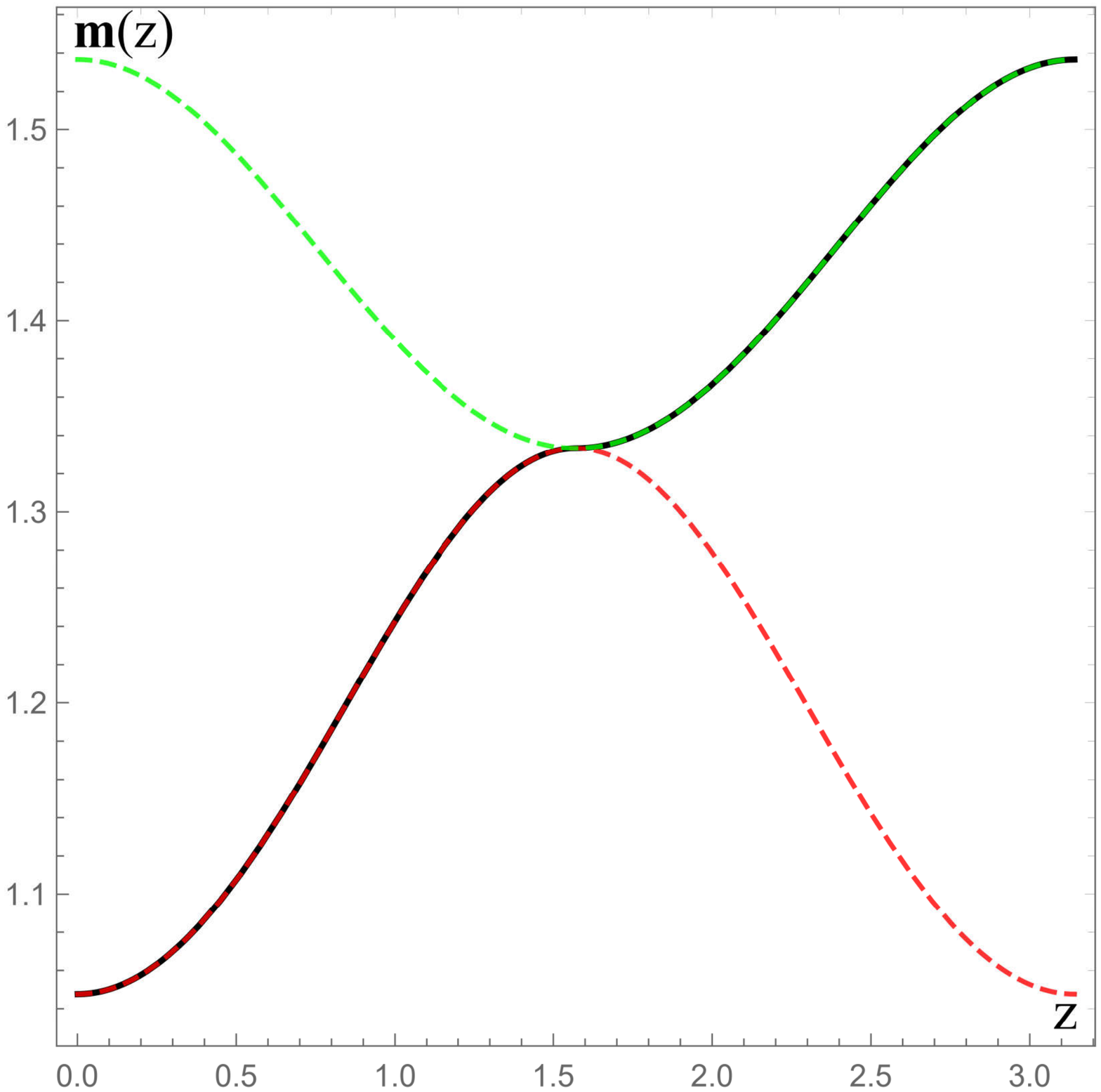}

(b) ${\bf m}'(0,\frac{\pi}{2})<0$\tabularnewline
\includegraphics[scale=0.22]{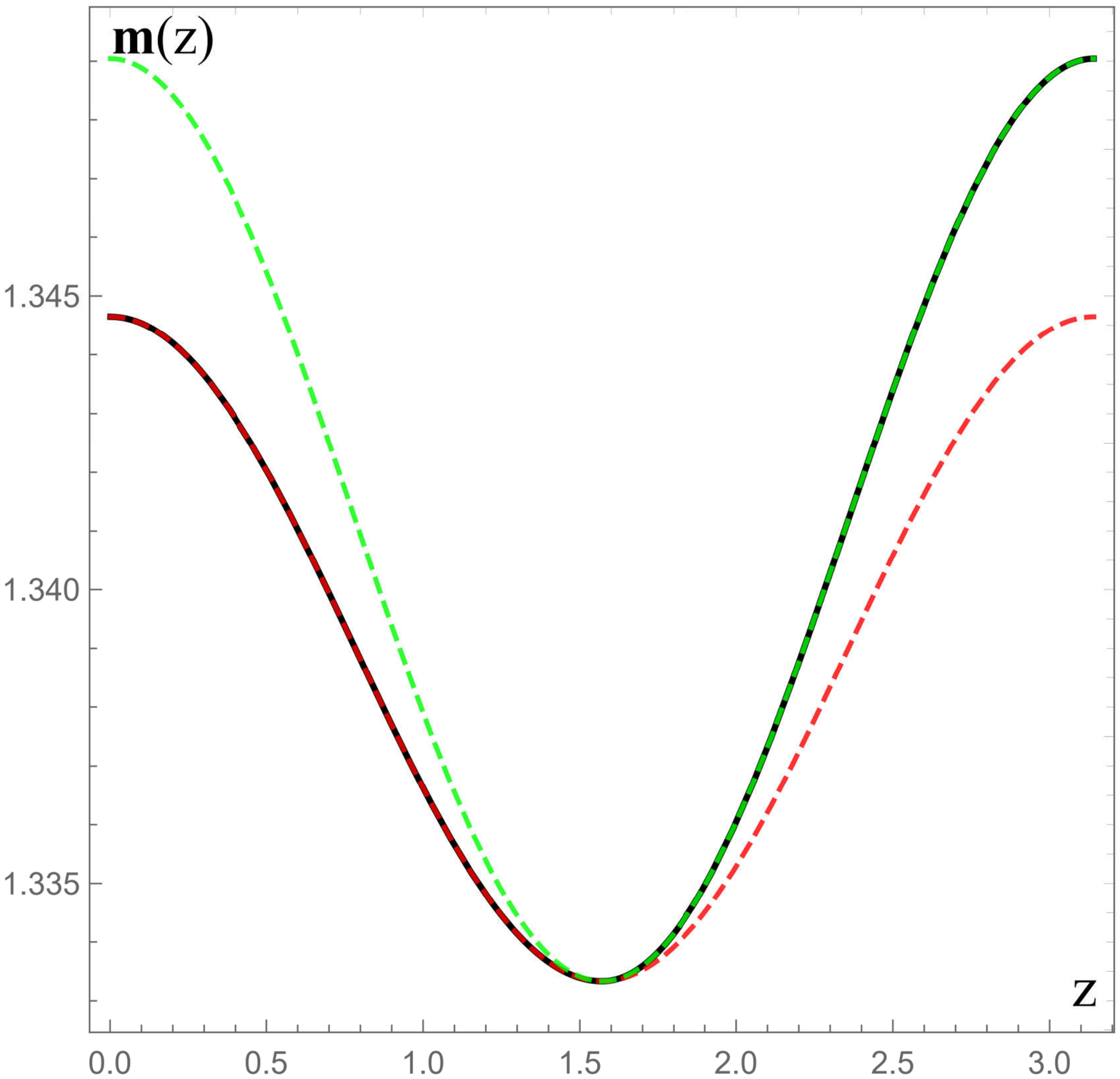}

(c) ${\bf m}'(0,\frac{\pi}{2})=0$, ${\bf m}''(0,\frac{\pi}{2})<0$. &
\includegraphics[scale=0.22]{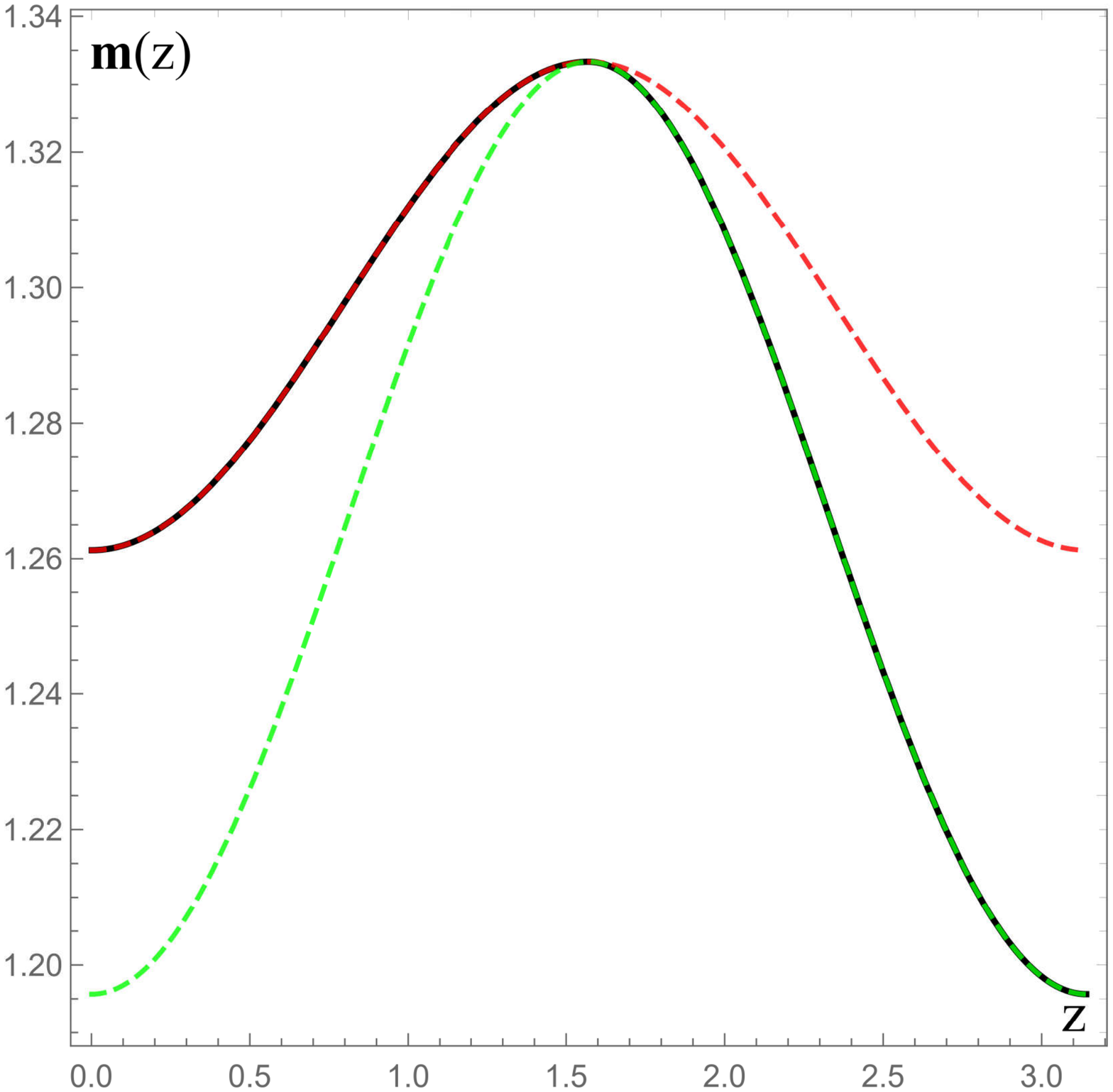}

(d) ${\bf m}'(0,\frac{\pi}{2})=0,$
${\bf m}''(0,\frac{\pi}{2})>0.$\tabularnewline
\end{tabular}
\par\end{centering}
\caption{\label{fig:RNFinalStateStable}The possible stable final states of the
RN-dS black hole under the perturbations when $(\hat{\Lambda}, Q)$ satisfy $P_2=2$. Here and in the next figure, the horizontal
axis is $z$ and the vertical axis is ${\bf m}(t,z)$.
 Note that the red and green lines are fitted  with ${\bf m}(z)=(1+Q^{2})\exp{(a\cos^2 z)}$ in which the parameter $a$ can be different  on the two hemispheres.
}
\end{figure}
As shown in Fig. \ref{fig:RNFinalStateStable}, on the critical line separating the stable and unstable regions, the evolution settles down eventually  to a stable lumpy black hole, depending on
the initial conditions. The mass and charge
densities are finite in the whole parameter space. The stable final states can be fitted very well by
\be
{\bf m}(z)=(1+Q^{2})e^{a\cos^2 z},
\ee
with the parameter $a$  describing the deviation from uniformity and being dependent on the initial conditions. Interestingly, we find the parameter $a$ can take different values on the two hemispheres,  which makes the horizon near $z=\pi/2$ not smooth in higher order derivatives. Though the physical origin for this  non-smoothness is unclear now, it is indeed permitted mathematically since the large $D$ effective equations are only of the first order in derivative.

Note that the non-smoothness at the equator is not in conflict with the static analysis in the previous section.
Since the quantization that $P_2$ is a positive integer is not a sufficient condition for  the smoothness of the solution at the equator, we can obtain lumpy black holes with different values of $P_1$ on the two hemispheres. In fact, the non-smoothness is not occasional for gravitational systems. For example,  let us consider a spherically symmetric black hole endowed with a spherical thin-shell located at some radius \cite{Cardoso:2013opa,Zhang:2014kna}. Then a junction condition must be imposed, in which the physical quantities such as the mass function is continuous but not analytical at the critical radius. Even worse, the discontinuity can appear in the gravitational systems, such as the studies of the shock waves \cite{Marolf:2011dj,Herzog:2016hob}.
%

\begin{figure}[h]
\begin{centering}
\begin{tabular}{>{\centering}p{6cm}>{\centering}p{6cm}}
\includegraphics[scale=0.23]{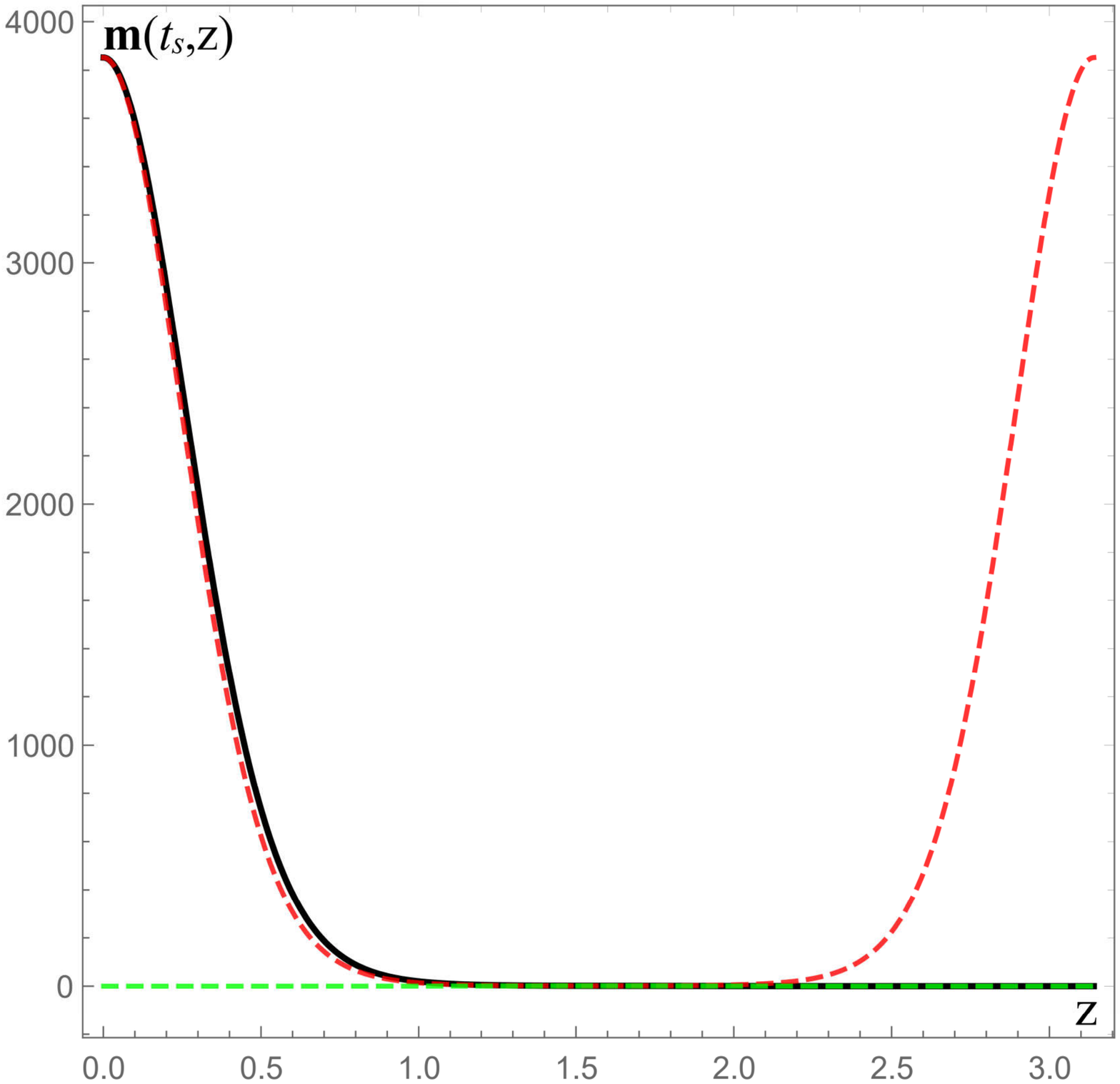}

(a) ${\bf m}'(0,\frac{\pi}{2})>0$ &

\includegraphics[scale=0.23]{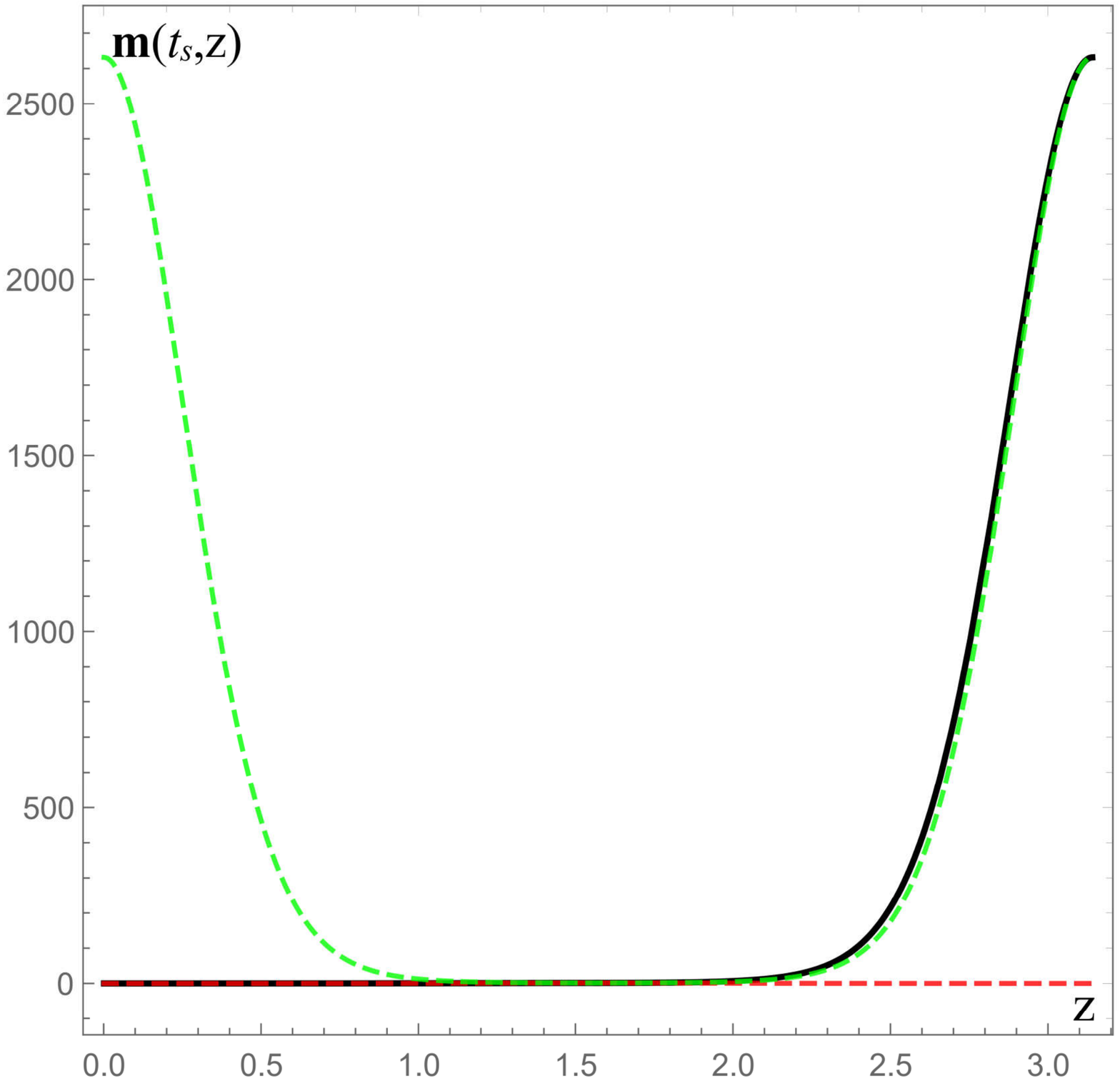}

(b) ${\bf m}'(0,\frac{\pi}{2})<0$\tabularnewline

\includegraphics[scale=0.23]{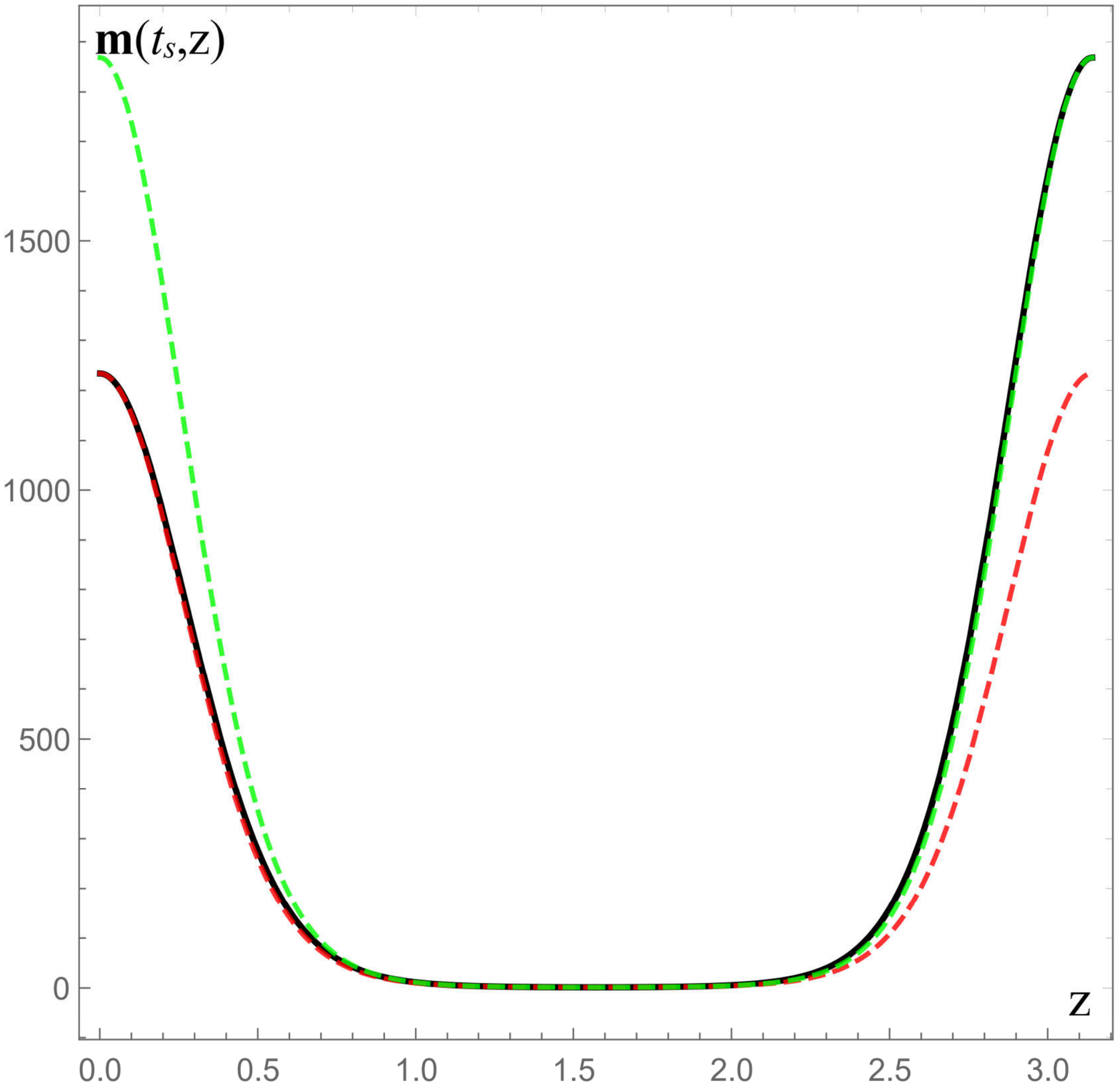}

(c) ${\bf m}'(0,\frac{\pi}{2})=0$, ${\bf m}''(0,\frac{\pi}{2})<0$. &

\includegraphics[scale=0.23]{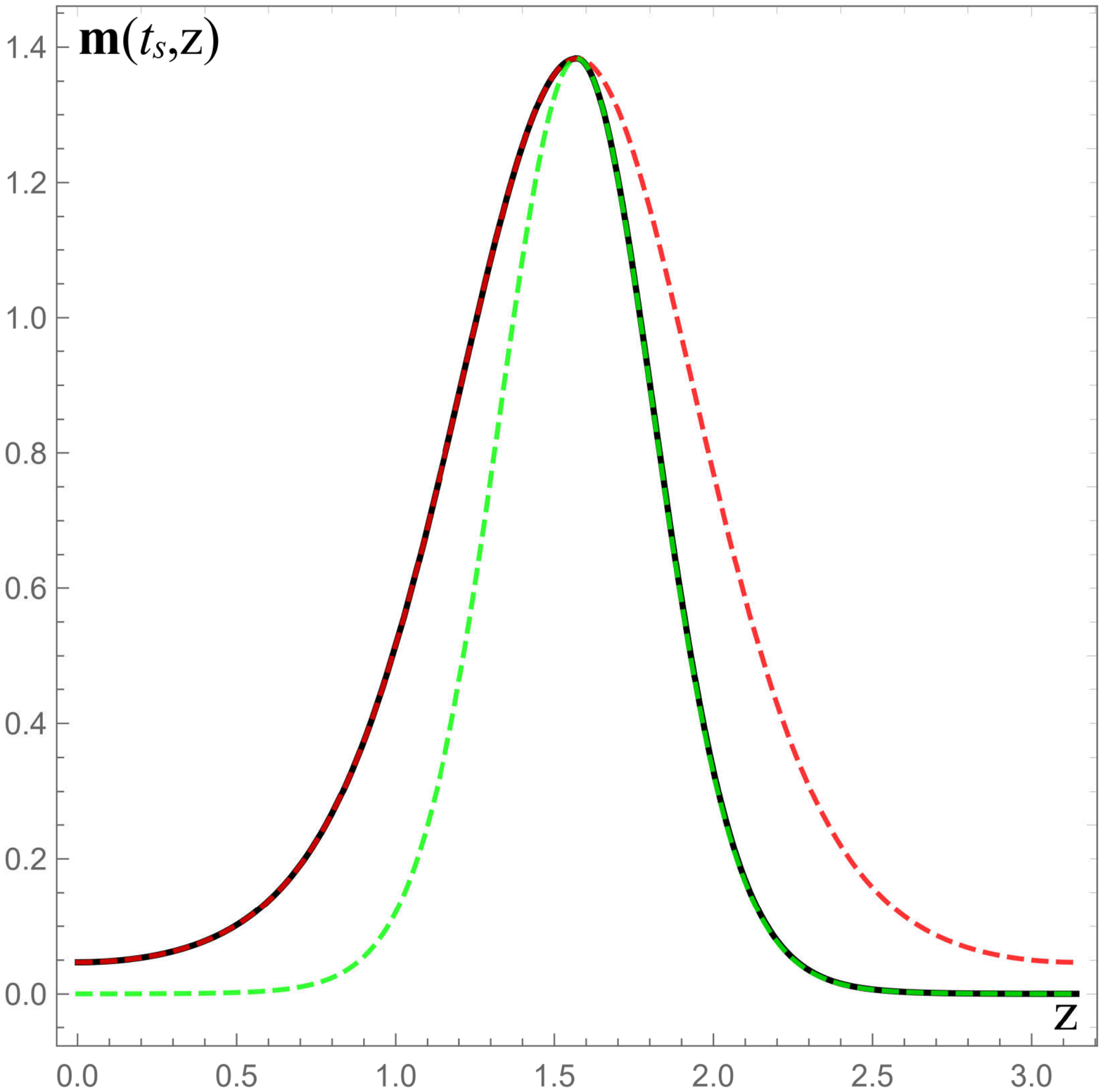}

(d) ${\bf m}'(0,\frac{\pi}{2})=0$, ${\bf m}''(0,\frac{\pi}{2})>0.$\tabularnewline
\end{tabular}
\par\end{centering}
\caption{\label{fig:RNFinalStateUnstable}The snap shots of the possible configurations of the unstable RN-dS black holes at some given instance $t_s$. Panel (a) describes the configuration that   the mass density locates mainly at the north pole, i.e. $z=0$. Similarly, in panel (b)  the mass density locates mainly at the south pole, i.e. $z=\pi$. In panel (c) almost the mass density occupy near both poles, so resembles a double spots.  In panel (d) the mass density focuses on the equator, i.e. $z=\pi/2$, so is complementary for panel (c). Note that the red and green lines are fitted lines with ${\bf m}(t,z)=(1+Q^{2})\exp({a e^{\omega t}\cos^2 z})$ in which the parameter $a$ can be different on the two hemispheres. }
\end{figure}

In the unstable region, i.e. for the parameters such that $P_2>2$, we find the black holes will not become stable but evolve with time. As shown in Fig.
\ref{fig:RNFinalStateUnstable}, the states at the late time could be classified into three kinds, with the mass function being focused  at the single pole, double poles or the equator of the sphere, depending on the initial conditions. Correspondingly,  these solutions resemble the fully localized black holes, which using the terminology from \cite{Dias1501,Dias1605} can be named as single spot, double spot or a black ring. In fact we can see that the stationary lumpy black holes that bifurcate from the onset of the instability  resemble these localized black holes as well.
As we will see the late time evolution can be fitted very well by
\be
{\bf m}(t,z)=(1+Q^{2})e^{a e^{\omega t}\cos^2 z},\label{massfunctionfit}
\ee
 with
 \be
 \omega=\frac{\sqrt{2(1+Q^2)^2\hat{\Lambda} +2Q^2-1}-1}{1+Q^2}.
\ee
Here the parameter $a$ can be different on the northern or southern hemisphere.
The mass density of the solutions grows very fast with time, thus when it becomes comparable with  $e^D$, the $1/D$ expansion will not be valid. So the large $D$ method can not really tell us the end state of the unstable RN-dS black hole. Nevertheless, the time evolution is suggestive, showing that the localized black spots and black ring solutions could exist as the end point of the ``$\Lambda$ instability'', and possibly also at finite $D$.

The effective equations (\ref{eq:effQ}), (\ref{eq:effM}) and (\ref{eq:effP}) allow us to study non-linearly the stabilities of the RN black holes in the asymptotically flat or the AdS spacetime by setting $\hat{\Lambda}=0$ or $\hat{\Lambda}\to-\hat{\Lambda}$. We find the RN black holes are always stable in these two cases, which is in accord with the linear analysis.

We further find an analytical solution for the large $D$ effective equations,
which has the form
\begin{align}
{\bf q}(t,z)=\, & \frac{Q}{1+Q^{2}} {\bf m}(t,z),\\
\quad {\bf p}(t,z)=\,&{\bf m}'(t,z)-\tan(z)\partial_{t}{\bf m}(t,z),\label{eq:RNdSDynamicP}\\
{\bf m}(t,z)=\, & c\exp\left[\sum_{\ell=2}^{\infty}\left(c_{\ell_+}e^{-i\omega_{\ell_+}t}+c_{\ell_-}
e^{-i\omega_{\ell_-}t}\right)\cos^{\ell}z\right].\label{eq:RNdSDynamicM}
\end{align}
 Here $c,c_{\ell\pm}$ are the integration constants and
\begin{equation}\label{fre:RNdS}
\omega_{\ell_\pm}=-i\frac{\ell-1\pm\sqrt{Q^{4}(\ell-1+\hat{\Lambda})\ell+2\hat{\Lambda} \ell Q^{2}-\ell(1-\hat{\Lambda} )+1}}{1+Q^{2}}.
\end{equation}
It turns out that this is exactly the quasinormal mode spectra for the scalar type gravitational perturbations of the RN-dS black holes found in \cite{Tanabe2015}. The modes with $\ell=0$ and 1 should be discarded since they violate the energy and momentum conservation, respectively.
It is easy to show that $\omega_{\ell-}$ always leads to the damping but $\omega_{\ell+}$ can lead the mode growing with time in the unstable region. The mode $\ell=2$ dominates the evolution since the growth rate decreases with $\ell$.\footnote{In the stable region, the decay rate increases with $\ell$, so the dominant mode is the one with $\ell=2$. This is reminiscent of the fact that in the ringdown phase of the coalescence of a binary black holes, the modes with $\ell=2$ are dominant.} In fact, the $\ell=2$ mode gives exactly the mass function \eqref{massfunctionfit}.

The analytical solution explains why the blue region in the left panel of Fig. \ref{fig:RNGBdSUnstableRegion} is unstable. For the parameters such that $P_2=N$, the mode with $\omega_{\ell=N}$ is static and the modes with $\omega_{\ell>N}$ decay off but the mode with $\omega_{\ell<N}$ can grow with time. The analytical solution matches well with the numerical evolutions in Fig. \ref{fig:RNFinalStateStable} and \ref{fig:RNFinalStateUnstable}, as long as $c_{\ell}$ take different values in the two regions $0\leq z<\pi/2$ and $\pi \geq z>\pi/2$. As we mentioned before, since the large $D$ effective equations are only of the first order in derivative, it is harmless to break the smoothness at $z=\pi/2$.

Although the non-smoothness at the equator is allowed in our study, we think it is a phenomenon emerging only at large $D$. For finite $D$, the dynamical equations is of the second order instead of the first order in derivatives, then the smoothness at the equator must be respected (at least up to the second order in derivatives). As a consequence, whether there appears the configuration with odd parity at the equator remains unclear. In particular, if in the evolution the mode with $\ell=2$ is still dominate then only solutions with even parity are allowable. But it might also be possible that the modes with odd $\ell$ can be excited and the modes with even $\ell$ are suppressed, then one may find  solutions with odd parity.

\subsection{GB-dS black hole}

Now we study the dynamical instability of the GB-dS black hole \footnote{More large $D$ studies on the GB black objects can be found in \cite{Chen2015,Chen1707,Chen1804, Chen1805,Saha:2018elg}.}. Since the whole discussion is  very similar so we will briefly present the result and omit the detail. At large $D$ the dynamics of the GB-dS black holes are encoded in the following effective equations \cite{Chen2017}
 \begin{align}
0= & \partial_{t}{\bf m}+\cot z\partial_{z}{\bf m}+\cot z\,{\bf p},\label{GBeq:effM}\\
0= & \partial_{t}{\bf p}-\frac{2(1+4\beta\hat{\Lambda}(1-\hat{\Lambda}))-(1+\beta)(1+2\beta(1-\hat{\Lambda})(2\hat{\Lambda}-1))}{(1+\beta)(1+2\beta\hat{\Lambda}(1-\hat{\Lambda}))}\cot z\partial_{z}{\bf p}\label{GBeq:effP}\\
 & +\left(1-\hat{\Lambda}-\frac{2\left[(1+4\beta\hat{\Lambda}(1-\hat{\Lambda}))-(1+\beta)(1+2\beta\hat{\Lambda}(1-\hat{\Lambda}) )\right]}{(1+\beta)(1+2\beta\hat{\Lambda}(1-\hat{\Lambda}))}\frac{{\bf p}}{{\bf m}}\cot z\right)\partial_{z} {\bf m}\nonumber \\
 & -\left(\frac{1-2\beta\hat{\Lambda}(1-\hat{\Lambda})-2\beta(1-\hat{\Lambda})^2 \cos2z}{(1+2\beta\hat{\Lambda}(1-\hat{\Lambda}))\sin^{2}z}-\frac{{\bf p}}{{\bf m}}\cot z-\frac{2(1+4\beta\hat{\Lambda}(1-\hat{\Lambda}))}{(1+\beta)(1+2\beta\hat{\Lambda}(1-\hat{\Lambda}))}\cot^{2}z\right){\bf p}.\nonumber
\end{align}
Note that the mass density ${\bf m}$ and the momentum density ${ \bf p}$ are related to the ones in \cite{Chen2017} by $m=(1-\hat{\Lambda})\,{\bf m}$, $p_z={ \bf p}$ and
$\beta$ is related to the GB parameter $\tilde{\alpha}$ by $\beta=(1-\hat{\Lambda})^{-1}\tilde{\alpha}$. Here we consider $\tilde{\alpha}>0$ and is of $\mc O(1)$ \footnote{The GB gravity involves the causality violation when quantum effects are considered unless the GB parameter is infinitesimal \cite{Camanho:2014apa}. However, our work is purely  classical and  the problem is evaded.} .
Similar to (\ref{eq:Deformed}), the static solutions of the large $D$ effective equation of the GB-dS black hole have the form ${\bf m}(z)=(1+\beta)e^{P_1(\cos z)^{P_2}}$  and ${\bf p}(z)={\bf m}'(z)$, where
$P_2$ is determined by $\beta,\hat{\Lambda}$ and given by
\be
P_2=\frac{2\beta(-3+\beta)\hat{\Lambda}^2+\hat{\Lambda}(1+(5-4\beta)\beta)+(1+\beta+2\beta^2)}
{4\beta(-1+\beta)\hat{\Lambda}^2-2\hat{\Lambda}\beta(-1+3\beta)+(1+\beta+2\beta^2)}.
\ee
At the edge of the instability, the parameters $(\beta,\hat{\Lambda})$ satisfy $P_2=\ell\geq2$ such that the solution is regular at $z\ge\pi/2$.
Moreover, these solutions have the same entropies in the microcanonical ensemble, so it is not possible to determine which one is preferred.

For the time evolution of the GB-dS black hole, the boundary and initial conditions are similar to those of the RN-dS black
hole. The threshold of the instability of the non-linear evolution is shown in the right panel of Fig. \ref{fig:RNGBdSUnstableRegion}. It coincides with the results from the linear analysis.
Note that only for a large enough cosmological constant in the presence of a positive GB parameter, the instability can be triggered.
Remarkably, we can see $\hat{\Lambda}\to1/\sqrt{3}\simeq0.577$  when $\beta\to\infty$. This is very close to that of the scalar type gravitational perturbations
shown in Fig. 3 in \cite{Cuyubamba2016} when $D=8$.
Thus although our analysis is done in the large $D$ limit, the result is suggestive for finite dimensions.

Similar to the RN-dS case shown in Fig. \ref{fig:RNFinalStateUnstable}, for the unstable GB-dS black holes the possible configurations at the late time can be classified into the ones resembling the  three kinds of fully localized black holes, with the mass density function being located about a single spot, double spots or a ring,
depending on the initial conditions.  On exactly the threshold of the instability,  there are three possible types of stable lumpy solutions,
just like the ones shown in Fig. \ref{fig:RNFinalStateStable}.
We also find an analytical solution of the same form as (\ref{eq:RNdSDynamicP},\ref{eq:RNdSDynamicM}) for the large $D$ effective equations, in which the  $\omega_{\ell\pm}$ are the quasinormal mode spectra for the scalar type gravitational perturbations of the GB-dS black holes in \cite{Chen2017}, i.e.
\be
\omega_{\ell\pm}=\frac{(-iU\pm \sqrt{V})}{(1+\beta)(1+2\beta(1-\hat{\Lambda}))},
\ee
where
\begin{align}
U= & (\ell-1)\left[1+2\beta(1-\hat{\Lambda}^{2}+\beta(\hat{\Lambda}-1)^{2})\right],\\
V= & -U^2-(1+\beta)(1+2\beta(1-\hat{\Lambda}))\ell[2\beta^2(1-\hat{\Lambda})(1-\hat{\Lambda}+\ell(2\hat{\Lambda}-1)) \\
   & +\beta(1+5\hat{\Lambda}-6\hat{\Lambda}^2+\ell(4\hat{\Lambda}^2-2\hat{\Lambda}-1))
   +1-\ell+\hat{\Lambda}].\nonumber
\end{align}
Besides, we study the non-linear evolutions of the GB black holes in the asymptotically flat and AdS spacetimes and find they are always stable.

\section{\label{sec:Conclusion}Summary and discussion}

In this work, we studied the non-linear dynamical instabilities  of the RN-dS and GB-dS black holes by employing the large $D$  method.  
The static solutions of the large $D$ effective equations had been previously constructed in \cite{Tanabe2015, Chen2017}, and were expected to exist at the zero modes of the instability. We found there are three branches for the lumpy black holes sharing the same horizon topology with the uniform solution. It turns out that the  lumpy solutions  have the same entropy with those of uniform solutions at the leading order in $1/D$.
It is not possible to determine which of these solutions is preferred from the viewpoint of thermodynamics at the leading order in $1/D$. Thus
we studied the time evolution of the black hole under the perturbations at the leading order in $1/D$. It turns out that the uniform solutions are preferred in the stable region, while the deformed solutions are preferred in the  unstable region and on the threshold.
The deformed solution on the threshold settles down, but is highly time-dependent in the unstable region. The solutions resemble the fully localized black holes with the mass density being localized at the north/south pole or the equator with $S^{D-2}$ or $S^1\times S^{D-3}$ topology, i.e., black spots or black ring, respectively. We found an analytical solution for the non-linear evolution, which matches well with the numerical simulation.
We also studied the non-linear evolutions of the RN and GB black holes in the asymptotically flat and AdS spacetimes and found they are always stable.

 Just as the entropy argument,   the time evolution of the black holes under the perturbations cannot determine  the  ultimate fate of these unstable  black holes  definitely to the leading order of $1/D$. One has to include the $1/D$ corrections to the large $D$ effective equations. It is expected that with the $1/D$ corrections the large $D$ effective equations would be of the second order in derivatives, like the ones for the large $D$ black strings \cite{Emparan:2015gva}. In the black string case, the second derivative terms are of comparable order. In contrast,  the second derivative terms  in the black hole case are smaller by an order of $1/D$. Therefore it is possible that the provided next-to-leading order $1/D$ corrections  would not change the picture qualitatively.

On the other hand, as stated in \cite{Andrade:2018nsz} the fact that the large $D$ effective equations of the spherical and spheroidal black hole systems are of the first order in derivative in polar angle $\theta$ stems from the assumption that $\theta=\mc O(1)$. As a consequence, some fine details of the dynamics that occur at shorter scales may be neglected. Ref. \cite{Andrade:2018nsz} found that the lumps on the non-uniform black branes can be viewed as localized black holes and their dynamics can be clearly captured by the large $D$ effective equations of the black branes. This approach can be taken as a  remedy, since it deals with the length scale $\sim \mc O(1/\sqrt{D})$ on the black brane, which corresponds to $\theta \sim \mc O(1/\sqrt{D})$ for the localized back holes. Thus, it is worth  applying this approach to the study of the RN-dS and GB-dS black holes. As a reference, using this approach the non-linear evolution of the Myers-Perry black holes agrees qualitatively well with the full numerical stimulation at finite $D$ \cite{Andrade:2019edf,Figueras:2017zwa,Bantilan:2019bvf}.

Despite some limitations, our study  suggests that there may be topology-changing transition for the RN-dS and GB-dS black holes in higher dimensions, which may indicate the violation of the Weak Cosmic Censorship. Therefore it would be interesting to explore the phase space at finite $D$ by numerically constructing the lumpy black holes and the possible localized black holes, just like the study for the black holes in AdS$_5\times$ S$^5$. The numerical study of AdS$_5$-Schwarzschild$\times$ S$^5$  shows that lumpy black holes are expected to connect to localized black hole solutions with the horizons $S^8$ or $S^4\times S^4$ based on the entropy argument \cite{Dias1501,Dias1605}. It would be possible that there is a similar transition between the lumpy black holes and the localized black holes  with $S^{D-2}$ or $S^1\times S^{D-3}$ topology for the RN-dS and GB-dS black holes. More recently, the authors in \cite{Bantilan:2019bvf} proposed a conjecture that the GL instability is the only mechanism that higher dimensional vacuum GR has to change the topology of a black hole horizon in dynamical spacetimes. Therefore it would be interesting to verify this conjecture in a broader setting.

It is worth pointing out that our study of the ``$\Lambda$ instability'' shows  similarities with the large $D$ studies of the instabilities of the black hole systems with spherical topology, such as the  ultraspinning instability \cite{Emparan2003,Dias2009} and bar-mode instability \cite{Shibata:2010wz} of the Myers-Perry black holes \cite{Suzuki:2015iha,Tanabe:2016opw}, and  the GL-like instability \cite{Banks98,Peet98,Hubeny02,Buchel1502} of the small  AdS$_D$-Schwarzschild$\times S^D$ black holes in supergravity \cite{Herzog2017}. We find that in all these cases, the large $D$ effective equations  are of the first order in derivative, there are lumpy solutions bifurcating from the onset of the instability, and the time evolution does not settle down to a stationary state (for large $D$ Myers-Perry black holes this has not been checked but we believe this is the case).

\section*{Acknowledgments}
We thank for the discussions with Yu Tian, Hong-Bao Zhang and Lin Chen. The work was in part supported by NSFC Grant No. 11335012,
No. 11325522, No. 11735001 and No. 11847241.  BC would like to thank the participants of the Third Symposium of BRICS-AGAC (Kazan, 29 August - 3 September, 2019) for stimulating discussions and comments.


\begin{thebibliography}{10}


\bibitem{Gregory1993} R. Gregory and R. Laflamme, ``Black strings and p-branes are unstable,'' Phys. Rev. Lett. {\bf70}
(1993) 2837, [hep-th/9301052]

\bibitem{Emparan2003}R. Emparan and R. C. Myers,``Instability of ultra-spinning black holes,'' JHEP {\bf09}, 025 (2003),
[hep-th/0308056]

\bibitem{Kol2006} B. Kol, ``The Phase transition between caged black holes and black strings: A Review,'' Phys. Rept. {\bf422},119 (2006), [hep-th/0411240].

\bibitem{Harmark2007} T. Harmark, V. Niarchos, and N. A. Obers, ``Instabilities of black strings and branes,'' Class. Quant. Grav. {\bf24}, R1 (2007), [hep-th/0701022].

\bibitem{Lehner2010} L. Lehner and F. Pretorius, ``Black Strings, Low Viscosity Fluids, and Violation of Cosmic Censorship,'' Phys. Rev. Lett. {\bf105}, 101102 (2010), [arXiv:1006.5960].

\bibitem{Konoplya2008}R. A. Konoplya and A. Zhidenko, ``Instability of
higher dimensional charged black holes in the de-Sitter world,'' Phys. Rev. Lett.
{\bf103} (2009) 161101, arXiv:0809.2822 {[}hep-th{]}.

\bibitem{Konoplya:2013sba}
  R.~A.~Konoplya and A.~Zhidenko,
  ``Instability of D-dimensional extremally charged Reissner-Nordstrom(-de Sitter) black holes: Extrapolation to arbitrary D,''
  Phys.\ Rev.\ D {\bf 89}, no. 2, 024011 (2014)
  [arXiv:1309.7667 [hep-th]].

\bibitem{Cuyubamba2016}M. A. Cuyubamba, R. A. Konoplya and A. Zhidenko,
``Quasinormal modes and a new instability of Einstein-Gauss-Bonnet black
holes in the de Sitter world,'' Phys.Rev. D {\bf93} (2016) no.10, 104053,
arXiv:1604.03604 {[}gr-qc{]}.

\bibitem{Zhang2015} C. Y. Zhang, S. J. Zhang, D. C. Zou, B. Wang, Charged scalar
    gravitational collapse in de Sitter spacetime, Phys.Rev. D{\bf 93} (2016) no.6, 064036, arXiv:1512.06472 [gr-qc].



\bibitem{Emparan2013}R. Emparan, R. Suzuki and K. Tanabe, ``The large
$D$ limit of General Relativity,'' JHEP {\bf1306}, 009 (2013), arXiv:1302.6382.

\bibitem{Emparan:2014aba}
  R.~Emparan, R.~Suzuki and K.~Tanabe,
  ``Decoupling and non-decoupling dynamics of large D black holes,''
  JHEP {\bf 1407}, 113 (2014)
  [arXiv:1406.1258 [hep-th]].

\bibitem{Emparan:2015hwa}
  R.~Emparan, T.~Shiromizu, R.~Suzuki, K.~Tanabe and T.~Tanaka,
  ``Effective theory of Black Holes in the 1/D expansion,''
  JHEP {\bf 1506}, 159 (2015)
  [arXiv:1504.06489 [hep-th]].

  \bibitem{Bhattacharyya:2015dva}
  S.~Bhattacharyya, A.~De, S.~Minwalla, R.~Mohan and A.~Saha,
  ``A membrane paradigm at large D,''
  JHEP {\bf 1604}, 076 (2016)
  [arXiv:1504.06613 [hep-th]].

  \bibitem{Suzuki:2015iha}
  R.~Suzuki and K.~Tanabe,
  ``Stationary black holes: Large $D$ analysis,''
  JHEP {\bf 1509}, 193 (2015)
  [arXiv:1505.01282 [hep-th]].

  \bibitem{Emparan:2015gva}
  R.~Emparan, R.~Suzuki and K.~Tanabe,
  ``Evolution and End Point of the Black String Instability: Large D Solution,''
  Phys.\ Rev.\ Lett.\  {\bf 115}, no. 9, 091102 (2015)
  [arXiv:1506.06772 [hep-th]].

\bibitem{Bhattacharyya:2015fdk}
  S.~Bhattacharyya, M.~Mandlik, S.~Minwalla and S.~Thakur,
  ``A Charged Membrane Paradigm at Large D,''
  JHEP {\bf 1604}, 128 (2016)
  [arXiv:1511.03432 [hep-th]].

\bibitem{Dandekar:2016fvw}
  Y.~Dandekar, A.~De, S.~Mazumdar, S.~Minwalla and A.~Saha,
  ``The large D black hole Membrane Paradigm at first subleading order,''
  JHEP {\bf 1612}, 113 (2016)
  [arXiv:1607.06475 [hep-th]].

 \bibitem{Bhattacharyya:2017hpj}
  S.~Bhattacharyya, P.~Biswas, B.~Chakrabarty, Y.~Dandekar and A.~Dinda,
  ``The large D black hole dynamics in AdS/dS backgrounds,''
  JHEP {\bf 1810}, 033 (2018)
  [arXiv:1704.06076 [hep-th]].

\bibitem{Bhattacharyya:2018szu}
  S.~Bhattacharyya, P.~Biswas and Y.~Dandekar,
  ``Black holes in presence of cosmological constant: second order in $ \frac{1}{D} $,''
  JHEP {\bf 1810}, 171 (2018)
  [arXiv:1805.00284 [hep-th]].

 \bibitem{Kundu:2018dvx}
  S.~Kundu and P.~Nandi,
  ``Large D gravity and charged membrane dynamics with nonzero cosmological constant,''
  JHEP {\bf 1812}, 034 (2018)
  [arXiv:1806.08515 [hep-th]].

\bibitem{Tanabe2015}K. Tanabe, ``Instability of de Sitter Reissner-Nordstrom
black hole in the $1/D$ expansion,'' Class. Quant. Grav. {\bf33} (2016) no.12,
125016, arXiv:1511.06059 {[}hep-th{]}.

\bibitem{Chen2017}B. Chen and P.-C. Li, ``Static Gauss-Bonnet Black Holes
at Large $D$,'' JHEP {\bf1705} (2017) 025, arXiv:1703.06381 {[}hep-th{]}.


\bibitem{Suzuki1506}R. Suzuki and K. Tanabe, `` Non-uniform black strings and the critical dimension in the
$1/D$ expansion," JHEP {\bf10} (2015) 107, [arXiv:1506.01890].


  \bibitem{Rozali1607} M. Rozali and A. Vincart-Emard, ``On Brane Instabilities in the Large $D$ Limit,'' JHEP {\bf 1608}, 166 (2016) [arXiv:1607.01747].

\bibitem{Emparan:2018bmi}
  R.~Emparan, R.~Luna, M.~Martinez, R.~Suzuki and K.~Tanabe,
  ``Phases and Stability of Non-Uniform Black Strings,''
  arXiv:1802.08191 [hep-th].


\bibitem{Cardoso:2013opa}
  V.~Cardoso, I.~P.~Carucci, P.~Pani and T.~P.~Sotiriou,
  ``Matter around Kerr black holes in scalar-tensor theories: scalarization and superradiant instability,''
  Phys.\ Rev.\ D {\bf 88}, 044056 (2013)
  [arXiv:1305.6936 [gr-qc]].

\bibitem{Zhang:2014kna}
  C.~Y.~Zhang, S.~J.~Zhang and B.~Wang,
  ``Superradiant instability of Kerr-de Sitter black holes in scalar-tensor theory,''
  JHEP {\bf 1408}, 011 (2014)
  [arXiv:1405.3811 [hep-th]].



\bibitem{Marolf:2011dj}
  D.~Marolf and A.~Ori,
  ``Outgoing gravitational shock-wave at the inner horizon: The late-time limit of black hole interiors,''
  Phys.\ Rev.\ D {\bf 86}, 124026 (2012)
  [arXiv:1109.5139 [gr-qc]].

\bibitem{Herzog:2016hob}
  C.~P.~Herzog, M.~Spillane and A.~Yarom,
  ``The holographic dual of a Riemann problem in a large number of dimensions,''
  JHEP {\bf 1608}, 120 (2016)
  [arXiv:1605.01404 [hep-th]].

\bibitem{Chen2015}B. Chen, Z.-Y. Fan, P. Li and W. Ye, ``Quasinormal
modes of Gauss-Bonnet black holes at large $D$,'' JHEP {\bf1601} (2016) 085,
arXiv:1511.08706 {[}hep-th{]}.




\bibitem{Chen1707}B. Chen, P.-C. Li and C.-Y. Zhang, ``Einstein-Gauss-Bonnet
Black Strings at Large $D$,'' JHEP {\bf1710} (2017) 123, arXiv:1707.09766 {[}hep-th{]},

\bibitem{Chen1804} B. Chen, P.-C. Li, Yu Tian and C.-Y. Zhang, ``Holographic Turbulence in Einstein-Gauss-Bonnet
Gravity at Large $D$,'' JHEP {\bf01} (2019)156 [arXiv:1804.05182].

\bibitem{Chen1805}B. Chen, P.-C. Li and C.-Y. Zhang, ``Einstein-Gauss-Bonnet
Black Rings at Large $D$,'' JHEP {\bf1807} (2018) 067 , arXiv:1805.03345 {[}hep-th{]}.

	
\bibitem{Saha:2018elg}
  A.~Saha,
  ``The large D Membrane Paradigm For Einstein-Gauss-Bonnet Gravity,''
  JHEP {\bf 1901}, 028 (2019)
  [arXiv:1806.05201 [hep-th]].


\bibitem{Camanho:2014apa}
  X.~O.~Camanho, J.~D.~Edelstein, J.~Maldacena and A.~Zhiboedov,
  ``Causality Constraints on Corrections to the Graviton Three-Point Coupling,''
  JHEP {\bf 1602}, 020 (2016)
  [arXiv:1407.5597 [hep-th]].

\bibitem{Andrade:2018nsz}
  T.~Andrade, R.~Emparan and D.~Licht,
  ``Rotating black holes and black bars at large D,''
  JHEP {\bf 1809}, 107 (2018)
  [arXiv:1807.01131 [hep-th]].

  \bibitem{Andrade:2019edf}
  T.~Andrade, R.~Emparan, D.~Licht and R.~Luna,
  ``Black hole collisions, instabilities, and cosmic censorship violation at large $D$,''
  arXiv:1908.03424 [hep-th].

  \bibitem{Figueras:2017zwa}
  P.~Figueras, M.~Kunesch, L.~Lehner and S.~Tunyasuvunakool,
  ``End Point of the Ultraspinning Instability and Violation of Cosmic Censorship,''
  Phys.\ Rev.\ Lett.\  {\bf 118}, no. 15, 151103 (2017)
  [arXiv:1702.01755 [hep-th]].

\bibitem{Bantilan:2019bvf}
  H.~Bantilan, P.~Figueras, M.~Kunesch and R.~Panosso Macedo,
  ``The End Point of Nonaxisymmetric Black Hole Instabilities in Higher Dimensions,''
  arXiv:1906.10696 [hep-th].

  \bibitem{Dias1501} O. J. C. Dias, J. E. Santos and B. Way, ``Lumpy $AdS_5\times S^5$ black holes and black belts,''
JHEP {\bf1504}, 060 (2015) [arXiv:1501.06574 [hep-th]].

\bibitem{Dias1605}O. J. C. Dias, J. E. Santos and B. Way, ``Localised $AdS_5\times S^5$ Black Holes,'' Phys. Rev.
Lett. {\bf117}, no. 15, 151101 (2016) [arXiv:1605.04911 [hep-th]].

  \bibitem{Dias2009}O. J. C. Dias, P. Figueras, R. Monteiro, J. E. Santos, and
R. Emparan, ``Instability and new phases of higher-dimensional rotating black holes,'' Phys. Rev. D{\bf80}, 111701 (2009), arXiv:0907.2248.

\bibitem{Shibata:2010wz}
  M.~Shibata and H.~Yoshino,
  ``Bar-mode instability of rapidly spinning black hole in higher dimensions: Numerical simulation in general relativity,''
  Phys.\ Rev.\ D {\bf 81}, 104035 (2010)
  [arXiv:1004.4970 [gr-qc]].

\bibitem{Tanabe:2016opw}
  K.~Tanabe,
  ``Charged rotating black holes at large D,''
  arXiv:1605.08854 [hep-th].


\bibitem{Banks98} T. Banks, M. R. Douglas, G. T. Horowitz and E. J. Martinec, ``AdS dynamics from
conformal field theory,'' [hep-th/9808016].

\bibitem{Peet98} A. W. Peet and S. F. Ross, ``Microcanonical phases of string theory on $AdS_m\times S^n$,''
JHEP {\bf9812}, 020 (1998) [hep-th/9810200].

\bibitem{Hubeny02} V. E. Hubeny and M. Rangamani, ``Unstable horizons,'' JHEP {\bf0205}, 027 (2002)
[hep-th/0202189].

\bibitem{Buchel1502}A. Buchel and L. Lehner, ``Small black holes in $AdS_5\times S^5$,'' Class. Quant. Grav. {\bf32}, no. 14, 145003 (2015) [arXiv:1502.01574 [hep-th]].

\bibitem{Herzog2017}C. P. Herzog and Y. Kim, ``The Large Dimension Limit
of a Small Black Hole Instability in Anti-de Sitter Space,'' JHEP {\bf1802}
(2018) 167, arXiv:1711.04865 {[}hep-th{]}.


\end{thebibliography}
\end{document}